\documentclass[twocolumn,aps,showpacs,prb,tightenlines,amsmath,amssymb]{revtex4}
\usepackage{graphicx}
\usepackage{amssymb}
\usepackage{dcolumn}
\usepackage{amsmath}
\usepackage{pifont}
\usepackage{bm}
\usepackage{colordvi}
\newcommand{\bgreek}[1]{\mbox{\boldmath$#1$\unboldmath}}

\begin{document}

\title{Electron spin diffusion in monolayer MoS$_2$}
\author{L. Wang}
\affiliation{Hefei National Laboratory for Physical Sciences at
Microscale and Department of Physics,
University of Science and Technology of China, Hefei,
Anhui, 230026, China}
\author{M. W. Wu}
\thanks{Author to whom correspondence should be addressed}
\email{mwwu@ustc.edu.cn.}
\affiliation{Hefei National Laboratory for Physical Sciences at
Microscale and Department of Physics, University of Science and
Technology of China, Hefei, Anhui, 230026, China}

\date{\today}

\begin{abstract}
Electron spin diffusion is investigated in monolayer MoS$_2$ in the absence of
external electric and magnetic fields. The electron-impurity scattering, which is shown to
play a negligible role in spin relaxation in time domain in this material, has a
marked effect on the in-plane spin diffusion due to the anisotropic spin precession
frequency in the spatial domain. With the electron-impurity and inter-valley
electron-phonon scatterings separately included in the scattering term, we study the intra- and
inter-valley diffusion processes of the in-plane spins by analytically solving the kinetic spin Bloch
equations. The intra-valley process is found to be dominant in the in-plane spin diffusion, in contrast to the case of
 spin relaxation in time domain, where the inter-valley process 
can be comparable to or even more
important than the intra-valley one. For the intra-valley process, we find that the
in-plane spin diffusion is suppressed with the increase of impurity density
but effectively enhanced by increasing electron density in both
the degenerate and nondegenerate limits. We also take into account the 
electron-electron Coulomb scattering in the intra-valley process. 
Interestingly, we find that in the nondegenerate limit, the intra-valley 
spin diffusion length 
presents an opposite trend in the electron density
dependence compared to the one with only electron-impurity scattering.
\end{abstract}


\maketitle
\section{INTRODUCTION}
Monolayer MoS$_2$ has attracted much attention due to its promising applications in
electronics,\cite{radisavljevic6,yzhang12,qhwang7,cg25}
optoelectronics,\cite{qhwang7,splendiani10,mak105,eda11,korn99,cao3,mak7,zeng7,xiao108,sallen86,lagarde} 
valleytronics\cite{cao3,mak7,zeng7,xiao108,sallen86,lagarde,tao,gwang,glazov2} and also
spintronics.\cite{zhu84,xiao108,chei,kadantsev152,ochoa,hshi,rostami,zahid,kormanyos,cappelluti,kosmider87,wangarXiv1,wangarXiv2}
For the application of spintronic devices, the suitable spin lifetime and spin
diffusion length are
required.\cite{nishikawa,kikkawa277,kikkawa397,martin67,voros94,tombros,datta,wunderlich,koo} 
This indicates the importance of the
investigations on the spin relaxation and spin diffusion in this material.

Very recently, spin relaxation has been studied in monolayer
MoS$_2$.\cite{ochoa,wangarXiv1,wangarXiv2} Wang and
Wu\cite{wangarXiv2} calculated the in-plane spin
relaxation time of electrons due to the D'yakonov-Perel'\cite{dp} (DP) and
Elliot-Yafet\cite{ey} (EY) mechanisms with the intra- and inter-valley processes
included. They pointed out that the DP mechanism, which results from the
inhomogeneous broadening\cite{wuning} together with any scattering
process, dominates the spin
relaxation. The inhomogeneous broadening is from the spin-orbit coupling (SOC) of the conduction band\cite{wangarXiv2}
\begin{equation}
{\bf \Omega}^{\mu}=[2\lambda_c\mu+\mu A_1k^2+A_2(k^3_x-3k_xk_y^2)]\hat{\bf z},
\label{soctime}
\end{equation}
where the $z$-axis is perpendicular to the monolayer MoS$_2$ plane; $\lambda_c$,
$A_1$ and $A_2$ are the strengths of the SOC; $\mu=1(-1)$ represents
the K(K$^{\prime}$) valley. The first term of the SOC, which is momentum independent, 
only induces the inter-valley DP spin
relaxation whereas the last two terms are momentum dependent, which lead to the intra- and inter-valley
spin relaxation processes. In addition, as only the last term (i.e., negligible anisotropic
cubic one) causes the DP spin relaxation with the electron-impurity
scattering, the electron-impurity
scattering is shown to play a marginal role in the spin relaxation.\cite{wangarXiv2}

In contrast to the spin relaxation in time domain, the inhomogeneous
broadening in spin diffusion for in-plane spins is 
determined by the spin precession frequency\cite{cheng101,weng}
\begin{equation}
 {\bgreek
  \omega}^{\mu}=\frac{m^*{\bf \Omega}^{\mu}}{\hbar^2k_x}=\frac{m^*}{\hbar^2}
\left[\frac{2\lambda_c\mu}{k_x}+\frac{\mu
A_1k^2}{k_x}+A_2(k^2_x-3k_y^2)\right]\hat{\bf z}
\label{socspatial}
\end{equation}
 when the spin
diffusion is along the $x$-axis. Here, $m^*$ stands for the
effective mass. Due to the existence of $k_x^{-1}$, all three terms 
become momentum dependent, which can induce the intra- and
inter-valley relaxations for in-plane spins along the diffusion. This is
different from the case of the spin relaxation in time domain as previously
mentioned. In addition, $k_x^{-1}$ also makes the first two terms (i.e., the
leading ones) anisotropic. This suggests that the electron-impurity
scattering may play an important role in the in-plane spin diffusion, which is
of great difference from the case of the spin relaxation in time domain in monolayer
MoS$_2$,\cite{wangarXiv2} but similar to the case of the spin diffusion in semiconductors\cite{cheng101,weng} and
single-layer graphene.\cite{peng84}

As for the out-of-plane spins, the spin diffusion length is infinite since the
spin precession frequency ${\bgreek \omega}^{\mu}$ [see Eq.~(\ref{socspatial})]
is along the out-of-plane direction. However,
this is not the case in the presence of an out-of-plane electric
field. Very recently, Bishnoi and
Ghosh\cite{bishnoi} investigated the out-of-plane spin diffusion with this 
electric field applied. They showed that the out-of-plane spins relax
during the spin diffusion since the out-of-plane electric field induces a Rashba SOC, which provides an
inhomogeneous broadening in the spatial domain for out-of-plane spins.\cite{cheng101} 
However, the Rashba SOC they used is incomplete according to the recent work by
Korm\'{a}nyos {\it et al.}.\cite{kormanyos2} 
In addition, the electron-electron Coulomb and electron-impurity
scatterings, which have been shown to
play an important role in spin diffusion in semiconductors\cite{cheng101,weng,cheng75,peng842} and
single-layer graphene,\cite{peng84} are absent
 in their work.\cite{bishnoi} Moreover, they also overlooked the
inter-valley electron-phonon scattering, which is of crucial importance in spin
relaxation in time domain.\cite{wangarXiv1,wangarXiv2} It is noted that in addition to the
out-of-plane electric field, an in-plane magnetic field can also lead to the out-of-plane spin relaxation along the 
spin diffusion.\cite{peng79,cheng101,weng} This is because the in-plane magnetic
field ({\bf B}) gives rise to a spin precession frequency in the spatial domain as
$m^*g\mu_B{\bf B}/(\hbar^2k_x)$, which provides an inhomogeneous broadening for out-of-plane
spins during the spin diffusion. 

In this work, we investigate the
electron spin diffusion in monolayer MoS$_2$ in the absence of the external
electric and magnetic fields. As the contribution of the spin-flip 
scattering due to the EY
mechanism is negligible,\cite{wangarXiv2,bishnoi} we only take into account the
spin conserving scattering. With the electron-impurity (inter-valley electron-phonon)
scattering included, the intra-valley (inter-valley) diffusion process for
in-plane spins is studied by analytically
solving the kinetic spin Bloch equations (KSBEs).\cite{wureview} 
We find that the intra-valley process dominates the in-plane spin diffusion, which 
is very different from the case
of the spin relaxation in time domain in monolayer MoS$_2$ where the inter-valley
process can be comparable to or even more important than the intra-valley one.\cite{wangarXiv2}
Moreover, it is shown that the in-plane spin diffusion length decreases with the
increase of the impurity density but increases with increasing electron
density in both the degenerate and nondegenerate limits. Very interestingly,
with the electron-electron Coulomb scattering further taken into account, 
the in-plane spin diffusion length shows an opposite electron density dependence
in the nondegenerate limit compared to the one with only the
  electron-impurity scattering.

This paper is organized as follows. In Sec.~II, we introduce our model and the
KSBEs. In Sec.~III, we investigate the in-plane spin diffusion 
by analytically solving the KSBEs. We summarize in Sec.~IV.

\section{MODEL AND KSBEs}
The effective Hamiltonian of the conduction band 
near the K(K$^{\prime}$) point in monolayer MoS$_2$ reads 
\begin{eqnarray}
H_{\rm eff}^{\mu}&=&\epsilon_{\mu{\bf k}}+{\bf \Omega}^{\mu}\cdot{\bgreek \sigma}/2\label{hamil}
\end{eqnarray}
according to the latest work by Wang and Wu.\cite{wangarXiv2} 
Here, $\epsilon_{\mu{\bf k}}=\hbar^2{\bf k}^2/(2m^*)$ with $m^*$ representing
the effective mass; ${\bf \Omega}^{\mu}$ is given in Eq.~(\ref{soctime}); ${\bgreek
  \sigma}$ are the Pauli matrices.

We then construct the microscopic KSBEs\cite{wureview} to investigate the 
electron spin diffusion in monolayer MoS$_2$. The KSBEs can be written as\cite{wureview} 
\begin{eqnarray}
  \partial_t{\rho}_{\mu{\bf k}}(x,t)&=&\left.\partial_t{\rho}_{\mu{\bf k}}(x,t)
\right|_{\rm
  coh}+\left.\partial_t{\rho}_{\mu{\bf k}}(x,t)\right|_{\rm
  scat}\nonumber\\
&&\mbox{}+\left.\partial_t{\rho}_{\mu{\bf k}}(x,t)\right|_{\rm dif},
\label{KSBE}
\end{eqnarray}  
with ${\rho}_{\mu{\bf k}}(x,t)$ being the density matrices of electrons at
position $x$ and time $t$. The diffusion terms are described as
\begin{equation}
\left.\partial_t{\rho}_{\mu{\bf k}}(x,t)\right|_{\rm dif}=-(\hbar k_x/m^*)\partial_x{\rho}_{\mu{\bf k}}(x,t)
\label{dif}
\end{equation}
by assuming that the spin diffusion is along the $x$-axis. 
The coherent terms $\left.\partial_t{\rho}_{\mu{\bf k}}(x,t)\right|_{\rm
  coh}$ can be found in Ref.~\onlinecite{yzhou}. As for the scattering terms
$\left.\partial_t{\rho}_{\mu{\bf k}}(x,t)\right|_{\rm scat}$, 
we neglect the spin-flip ones due to the EY mechanism since
the contribution of the EY mechanism is negligible.\cite{wangarXiv2,bishnoi}
Here, we only include the spin conserving terms, i.e, the electron-electron Coulomb, 
electron-impurity, intra-valley electron-acoustic phonon, electron-optical
phonon, and also the inter-valley electron-phonon\cite{kaasbjerg,wangarXiv1} (electron-KTA, -KLA,
-KTO, and -KLO) scatterings. Here, KTA, KLA, KTO and KLO correspond to the
transverse acoustic, longitudinal acoustic, transverse optical and longitudinal
optical phonon modes at the K point, respectively. The detailed expressions of
the above scattering terms are given in
Ref.~\onlinecite{yzhou} and the scattering matrix
elements are given in Ref.~\onlinecite{wangarXiv1}.

\section{SPIN DIFFUSION}
\subsection{Intra-valley process}
We first focus on the intra-valley diffusion process for in-plane spins by simplifying the
KSBEs with only the electron-impurity scattering included in the
scattering term. By taking the steady-state condition, 
the Fourier components of the density matrix with respect to
$\theta_{\bf k}$ are given by\cite{cubic}
\begin{eqnarray} 
&&\hspace{-1cm}\frac{\hbar
  k}{2m^*}\partial_x[\rho^{l+1}_{k}(x)+\rho^{l-1}_{k}(x)]\nonumber\\
&&\mbox{}=-\frac{i}{2\hbar}[(2\lambda_c+A_1k^2)\sigma_z,\rho^{l}_{k}(x)]
-\rho^{l}_{k}(x)/\tau_i^l,
\end{eqnarray}
where the valley index $\mu$ is neglected for the intra-valley process. 
Here, $[A,B]\equiv AB-BA$ is the commutator; $\rho^l_{{k}}(x)=\int_0^{2\pi}d\theta_{{\bf k}}\rho_{{\bf
    k}}(x)e^{-il\theta_{{\bf k}}}/(2\pi)$;
$1/\tau^l_i=N_im^*\int_0^{2\pi} d\theta_{\bf k}
U_{\bf q}^2(1-\cos l\theta_{\bf k})/(2\pi\hbar^3)$ stands for the $l$th-order momentum scattering
rate with $|{\bf q}|=\sqrt{2k^2(1-\cos\theta_{\bf k})}$. $N_i$ and $U_{\bf q}^2$ represent the impurity density and
electron-impurity scattering matrix element, respectively. It is noted that
  in recent experiments, the mobilities at room temperature are reported to
  be of the order of $0.1$-$10\ $cm$^2$V$^{-1}$s$^{-1}$ in monolayer MoS$_2$ on
  SiO$_2$ substrate.\cite{radisavljevic6,novoselov102,ayari101} By fitting to
  the mobilities in these experiments, one obtains the impurity densities 
(e.g., $N_i=4.4\times 10^{13}\ $cm$^{-2}$ corresponding to
  the mobility $10\ $cm$^2$V$^{-1}$s$^{-1}$).\cite{mobility} With these impurity
  densities, the electron-impurity scattering is in the strong scattering limit
  (in time domain), i.e., $\langle |2\lambda_c+A_1k^2|\rangle\tau^1_i/\hbar$ (e.g., $\sim 10^{-2}$ for
  $N_i=4.4\times 10^{13}\ $cm$^{-2}$) $\ll 1$, with
$\langle ... \rangle$ denoting the ensemble average. Then, 
one can only keep the lowest two orders of $|l|$ (i.e., 0,~1) and
obtain\cite{peng79,peng84,sun151} 
\begin{eqnarray}
\partial_x^2\rho^0_k(x)&=&-[\sigma_z,[\sigma_z,\rho^0_k(x)]]c_1^2/2+ic_2[\sigma_z,\rho^0_k(x)],
\end{eqnarray}
with $c_1=(2\lambda_c+A_1k^2)m^*/(\hbar^2 k)$ and
$c_2=(2\lambda_c+A_1k^2){m^*}^2/(\hbar^3 k^2\tau_i^1)$. By solving this equation under the boundary
conditions, the steady-state spin vector ${\bf S}_k(x)={\rm Tr}[\rho_k^0(x){\bgreek
  \sigma}]$ can be obtained. 

Since the system is isotropic for in-plane spins, we choose the injected spin
polarization along the $x$-axis without losing generality. With the boundary condition ${\bf S}_k(0)=(S_k(0),0,0)$ and ${\bf
  S}_k(+\infty)=0$, we have
\begin{eqnarray}
S_k^x(x)=S_k(0)\cos(\omega^{\rm in}_{\rm intra}x)e^{-x/l^{\rm in}_{\rm intra}},
\label{sy}
\end{eqnarray}  
with
\begin{eqnarray}
\hspace{-0.6cm}\omega^{\rm in}_{\rm
  intra}&=&\sqrt{\sqrt{c_1^4+c_2^2}+c_1^2},\\
\hspace{-0.6cm}1/l^{\rm in}_{\rm intra}&=&\sqrt{\sqrt{c_1^4+c_2^2}-c_1^2}.\label{lxintra}
\end{eqnarray}

\subsection{Inter-valley process}
We then turn to investigate the inter-valley spin diffusion process by
analytically solving the KSBEs with the inter-valley electron-phonon scattering
included in the scattering term. Similar to the intra-valley process, one obtains 
\begin{eqnarray}
\frac{\hbar k}{2m^*}\partial_x[\rho^{l+1}_{\mu k}(x)+\rho^{l-1}_{\mu
  k}(x)]&=&-\frac{i(2\lambda_c+A_1k^2)\mu}{2\hbar}\nonumber\\
&&\hspace{-4cm}\mbox{}\times[\sigma_z,\rho^{l}_{\mu
  k}(x)]-[{\rho^{l}_{\mu k}(x)}/{\tau_v^0}-{\rho^{l}_{-\mu
    k}(x)}/{\tau_v^l}]
\end{eqnarray}
based on the elastic scattering approximation.\cite{zhang112,lwang87,wangarXiv1}
Here, $1/\tau^l_v=(2N_{\rm ph}+1)m^*\int_0^{2\pi}
d\theta_{\bf k}M_{\bf q}^2\cos l\theta_{\bf k}/(2\pi\hbar^3)$ is the $l$th-order inter-valley momentum
scattering rate with $N_{\rm ph}$ and $M_{\bf q}^2$ being the
phonon number and the inter-valley electron-phonon scattering matrix element,
respectively. It is noted that differing from the electron-impurity
  scattering, the inter-valley electron-phonon scattering is in the weak
scattering limit (in time domain) according to the recent work by Wang and Wu,\cite{wangarXiv1} i.e., $\langle
|2\lambda_c+A_1k^2|\rangle|\tau^{0,1}_v|/\hbar\gg 1$. This 
indicates that high orders of $\rho_{\mu k}^l$ may be relevant in 
the inter-valley process. For simplicity, we first retain the lowest two orders
of $|l|$ (i.e., 0,~1) and obtain
\begin{eqnarray}
\partial_x^2\rho^0_{\mu k}(x)&=&-[\sigma_z,[\sigma_z,\rho^0_{\mu
  k}(x)]]c_1^2/2\nonumber\\
&&\hspace{-1cm}\mbox{}+ic_3\mu[\sigma_z,2\rho^0_{\mu k}(x)-\rho^0_{-\mu
k}(x)c_4],
\end{eqnarray}
with $c_3=(2\lambda_c+A_1k^2){m^*}^2/(\hbar^3 k^2\tau_v^0)$ and
$c_4=1-\tau^0_v/\tau^1_v$. From this equation together with certain boundary conditions, we can obtain the
steady-state spin vector ${\bf S}_{\mu k}(x)={\rm Tr}[\rho_{\mu k}^0(x){\bgreek
  \sigma}]$. Similar to the intra-valley process, we also set the initial spin
polarization along the $x$-axis. Under the boundary condition that ${\bf S}_{\mu k}(0)=(S_k(0),0,0)$ and ${\bf
  S}_{\mu k}(+\infty)=0$, we have
\begin{eqnarray}
\sum_{\mu}S_{\mu k}^x(x)=2S_k(0)\cos(\omega^{\rm in}_{\rm inter}x)e^{-x/l^{\rm in}_{\rm inter}},
\label{syinter}
\end{eqnarray}  
where 
\begin{eqnarray}
\hspace{-0.4cm}\omega^{\rm in}_{\rm inter}&=&\sqrt{2}c_1,\\
\hspace{-0.4cm}1/{l^{\rm in}_{\rm
  inter}}&=&{c_3}\sqrt{4-c_4^2}/({\sqrt{2}c_1}).\label{lxinter}
\end{eqnarray}
With higher orders ($|l|>1$) of $\rho_{\mu k}^l$ included, we find
that the inter-valley spin diffusion length $l^{\rm in}_{\rm inter}$ varies slightly 
 (not shown).\cite{highorder}  

\subsection{Discussion and results}
In the calculation, the
effective mass $m^*=0.38m_0$ with $m_0$ being the
free electron mass;\cite{wangarXiv2} the coefficients of the SOC $\lambda_c=1.5\
$meV [Ref.~\onlinecite{kormanyos}] and $A_1=417.94\ $meV\ \r{A}$^2$.\cite{wangarXiv2} The parameters related to the 
scattering matrix elements are given in Ref.~\onlinecite{wangarXiv1}. With these
parameters, we calculate the intra- and inter-valley spin diffusion lengths in
both the degenerate and nondegenerate limits.

In the degenerate limit, the intra- and inter-valley spin diffusion lengths are
given by $l^{\rm in}_{\rm intra}(k_F)$ and $l^{\rm in}_{\rm inter}(k_F)$ according to
Eqs.~(\ref{lxintra}) and (\ref{lxinter}), respectively. Here,
$k_F$ is the Fermi wavevector. To compare the relative importance between the
intra- and inter-valley processes in the spin diffusion, we calculate
\begin{eqnarray}
\frac{l^{\rm in}_{\rm intra}(k_F)}{l^{\rm in}_{\rm inter}(k_F)}=\frac{c_3}{\sqrt{2}c_1}\sqrt{4-c_4^2}/\sqrt{\sqrt{c_1^4+c_2^2}-c_1^2}.
\end{eqnarray}
As the electron-impurity scattering is in
the strong scattering limit in time domain (i.e., $c_1^2\ll c_2$) and the
 inter-valley
electron-phonon scattering is in the weak scattering limit in time domain (i.e., $c_1^2\gg
c_3$), we have $l^{\rm in}_{\rm intra}(k_F)\ll l^{\rm in}_{\rm inter}(k_F)$.  
This indicates that the intra-valley process dominates the in-plane spin diffusion, which
is very different from the case of the spin relaxation in time domain in
monolayer MoS$_2$ where
the inter-valley process is shown to be comparable or even more important than the
intra-valley one.\cite{wangarXiv2} In the following, we only calculate the spin diffusion
length due to the intra-valley process $l^{\rm in}_{\rm intra}(k_F)$ according to Eq.~(\ref{lxintra}).

\begin{figure}[bth]
\centering
\includegraphics[width=8.5cm]{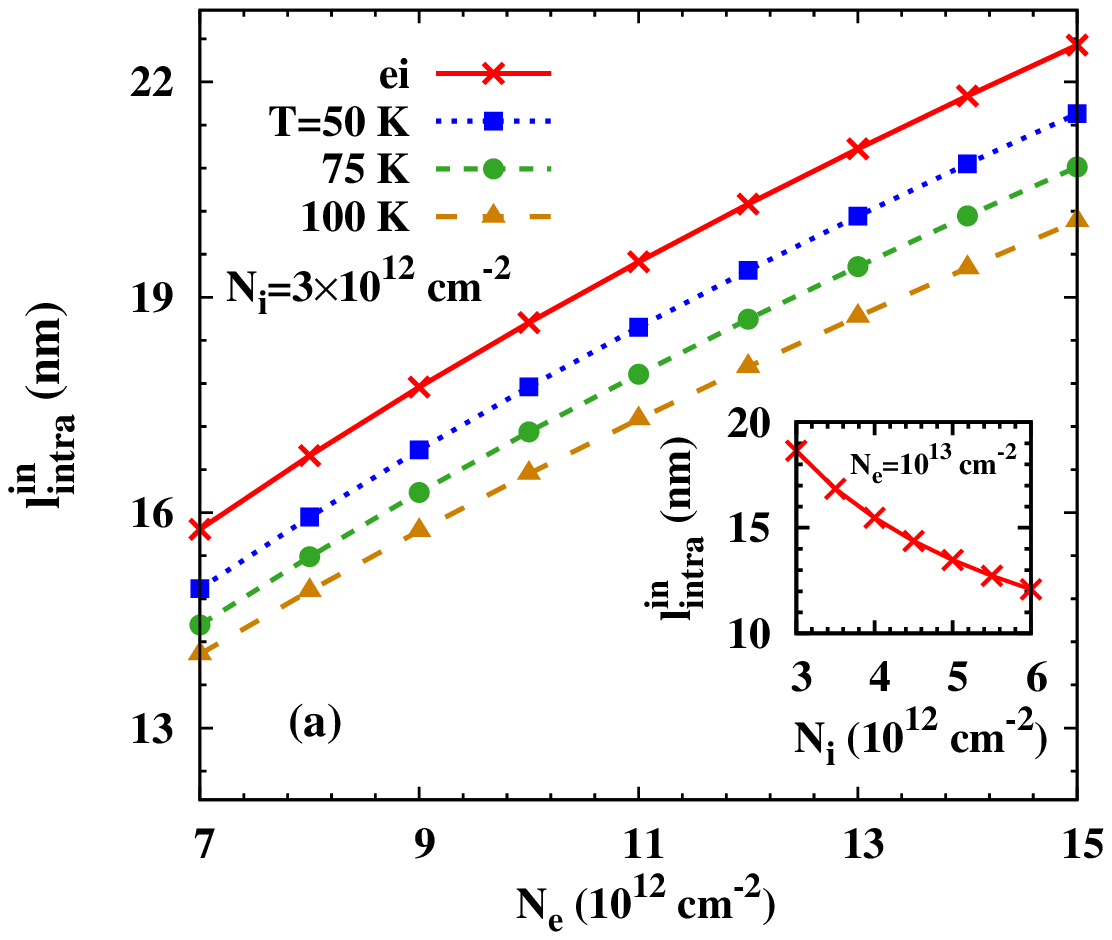}
\includegraphics[width=8.5cm]{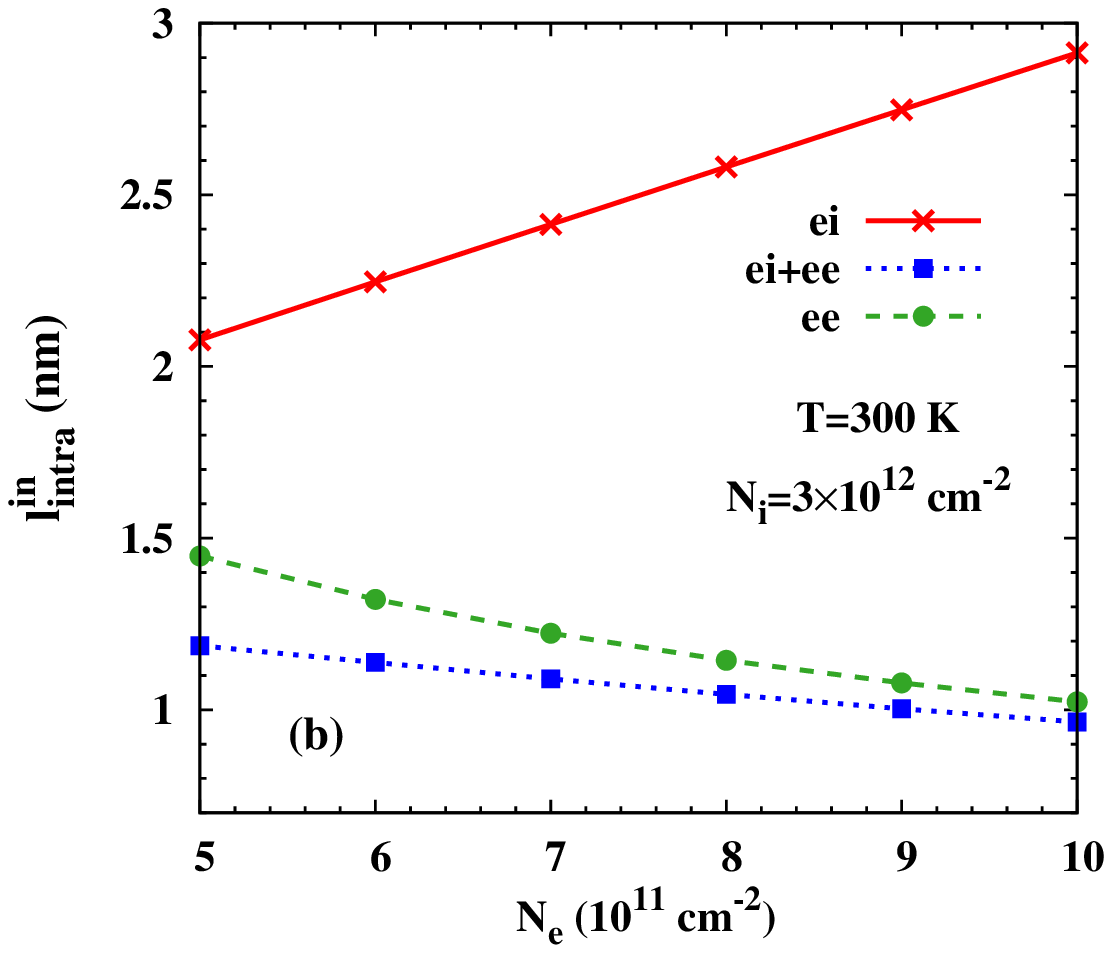}
\caption{(Color online) (a) Intra-valley spin diffusion length for in-plane spins $l^{\rm in}_{\rm intra}$ 
as function of the electron density $N_e$ in the degenerate limit. The curve with $\times$ stands for the calculation
from Eq.~(\ref{lxintra}) with only the electron-impurity
  scattering included whereas the curves with $\blacksquare$, $\bullet$ and $\blacktriangle$ represent
  the calculation based on Eq.~(\ref{lxintra}) with the effective momentum
  scattering rate $1/\tau^*$ in place of $1/\tau^1_i$ at 
  temperature $T=50$, $75$ and $100\ $K, respectively. The impurity density
  $N_i=3\times 10^{12}\ $cm$^{-2}$. In addition, $l^{\rm in}_{\rm intra}$ due to the
  electron-impurity scattering as function of the impurity density $N_i$ with
  $N_e=10^{13}\ $cm$^{-2}$ is shown in the inset; (b) $l^{\rm in}_{\rm intra}$
  as  function of $N_e$ in the nondegenerate limit. The curves with $\times$,
  $\bullet$ and $\blacksquare$ stand for the calculation from
  Eq.~(\ref{lxintra}) with the momentum scattering rate being $1/\tau^1_i$, $1/\tau_{\rm ee}$ and $1/\tau^*$,
  respectively. $N_i=3\times 10^{12}\ $cm$^{-2}$ and $T=300\ $K.}
\label{fig1}
\end{figure}

In Fig.~\ref{fig1}(a), we plot the electron density dependence of the
in-plane spin diffusion length. The impurity density is taken to be $N_i=3\times 10^{12}\
$cm$^{-2}$ so that the electron-impurity scattering is in the strong scattering
limit in time domain. The corresponding mobility is of the order of $10^2\
$cm$^2$V$^{-1}$s$^{-1}$, which is about $1$-$2$ orders of magnitude larger than
those reported in the experiments.\cite{radisavljevic6,novoselov102,ayari101} 
We find that the spin diffusion length due to the
electron-impurity scattering (curve with $\times$) increases with the increase of the electron
density. This can be understood as follows. The intra-valley spin diffusion
length is approximately given by 
\begin{eqnarray}
l^{\rm in}_{\rm intra}(k_F)=1/\sqrt{c_2}\label{lxapp}
\end{eqnarray}
according to Eq.~(\ref{lxintra}) under the condition that $c_1^2\ll c_2$ as
previously mentioned. When the electron density increases,
the Fermi wavevector $k_F$ increases whereas the electron-impurity scattering
rate $1/\tau_{i}^1$ decreases in the degenerate limit,\cite{yzhou} leading 
to the decrease of $c_2$ and therefore the increase of the spin diffusion
length. In addition to the
electron-impurity scattering, the electron-electron
Coulomb scattering is also taken into account whereas the intra-valley electron-phonon
scattering is neglected due to the negligible contribution.\cite{wangarXiv2}
As done in the spin relaxation in time domain,\cite{wureview,glazov} we calculate the in-plane spin diffusion
length according to Eq.~(\ref{lxintra}) with the effective momentum scattering
rate $1/\tau^*=1/\tau_i^1+1/\tau_{\rm ee}$ replacing the one due to the
electron-impurity scattering $1/\tau_i^1$. Here, $\tau_{\rm
  ee}^{-1}=(\pi/4)\ln(E_F/k_BT)k_B^2T^2/(\hbar E_F)$ represents the 
momentum scattering rate due to the electron-electron Coulomb
scattering with $E_F$ and $k_B$ being the Fermi energy and Boltzmann constant, respectively.\cite{glazov}  
The results at $T=50$, $75$ and $100\ $K are shown in Fig.~\ref{fig1}(a). By comparing
these results with the one calculated with only the electron-impurity scattering
(curve with $\times$), we find that that the electron-impurity scattering plays a
leading role in the in-plane spin diffusion. This is very different from the case of the spin relaxation
in time domain in monolayer MoS$_2$ where the contribution of the electron-impurity scattering
is marginal.\cite{wangarXiv2} Moreover, we find that the intra-valley spin
diffusion length decreases with the increase of the temperature when the
electron density remains unchanged. This is because the electron-impurity
scattering is insensitive to the temperature in the degenerate
limit\cite{yzhou,lwang87} whereas the electron-electron
Coulomb scattering increases with increasing temperature in the degenerate
limit,\cite{glazov} leading to the decrease of the spin diffusion length as the
temperature increases [see Eq.~(\ref{lxapp})].

In addition, we also investigate the
impurity density dependence of the in-plane spin diffusion length with the electron
density $N_e=10^{13}\ $cm$^{-2}$ as shown in the inset 
of Fig.~\ref{fig1}(a), with only the electron-impurity scattering included.
It is seen that the spin diffusion length decreases with the increase of
the impurity density, which can be easily understood from
Eq.~(\ref{lxapp}). 

Then we turn to the nondegenerate limit. The intra- and inter-valley spin
diffusion lengths are calculated by $l^{\rm in}_{\rm intra}(k_T)$ and $l^{\rm in}_{\rm inter}(k_T)$ according to
Eqs.~(\ref{lxintra}) and (\ref{lxinter}), respectively. Here,
$k_T=\sqrt{2m^*k_BT}/\hbar$ stands for the ``thermal'' wavevector.\cite{glazov}
Similar to the degenerate-limit case, the in-plane spin diffusion in the
nondegenerate limit is also dominated by the intra-valley process. With the
impurity density $N_i=3\times 10^{12}\ $cm$^{-2}$, we calculate the electron density dependence of $l^{\rm in}_{\rm
  intra}(k_T)$ due to the electron-impurity scattering and the result is shown in
Fig.~\ref{fig1}(b). We find that the intra-valley spin
diffusion length increases with the increase of the electron density, which
results from the decrease of $1/\tau^1_i$ due to the increase of the screening in
the nondegenerate limit.\cite{wangarXiv1} With the effective momentum scattering rate
$1/\tau^*=1/\tau_i^1+1/\tau_{\rm ee}$ instead of $1/\tau_i^1$, we also take into account the
electron-electron Coulomb scattering . It is noted that differing from the case
of the degenerate limit, $1/\tau_{\rm
  ee}=35.7e^4N_e/(\hbar\kappa^2k_BT)$ in the nondegenerate limit with $\kappa$
being the relative static dielectric constant.\cite{glazov} It is seen that the
spins relax faster along the spin diffusion with increasing electron density,
which is opposite to the situation with only the electron-impurity
scattering included. This can be understood as follows. 
As the electron density increases, the enhancement of $1/\tau_{\rm
  ee}$ leads to the decrease of the intra-valley spin diffusion length due to the
electron-electron Coulomb scattering (curve with $\bullet$), which suppresses
the increase of $l^{\rm in}_{\rm intra}(k_T)$ contributed by the electron-impurity
scattering.

It is noted that the feature of the spin diffusion is similar to the one
 of the spin relaxation in the weak scattering limit in time 
domain.\cite{wureview} Specifically, the
 intra-valley spin diffusion length decreases with increasing the
scattering. This is because the spin precession frequency in the
 spatial domain is proportional to $k_x^{-1}$,\cite{magnetic} which gives 
strong inhomogeneous broadening along the spin diffusion. 
Similar behavior has also been shown in the case of
 spin diffusion in symmetric silicon quantum wells under an in-plane
magnetic field\cite{peng79} since the spin precession frequency in the
  spatial domain has similar momentum dependence.\cite{magnetic}

\section{Summary}
In conclusion, we have investigated the electron spin diffusion in monolayer
MoS$_2$ in the absence of the external electric and magnetic fields. 
The anisotropic spin precession frequency in the spatial domain leads to a 
marked contribution of the
electron-impurity scattering in the in-plane spin diffusion. This is of great difference
from the spin relaxation in time domain in monolayer MoS$_2$ where the
contribution of the electron-impurity scattering is negligible. With 
the electron-impurity and inter-valley electron-phonon scatterings separately included in
the KSBEs, we analytically study the intra- and inter-valley diffusions for
in-plane spins. The
intra-valley process is shown to play a dominant role in the in-plane spin
diffusion, which is different
from the case of the spin relaxation in time domain in this material where the
inter-valley process can be comparable to or even more important than the
intra-valley one. In the dominant intra-valley process, it is shown that the in-plane
spin diffusion is suppressed by increasing impurity density but enhanced with
the increase of the electron density in both the
degenerate and nondegenerate limits. Interestingly, with the electron-electron Coulomb
scattering further included, a decrease of the spin diffusion length is observed
as the electron density increases in the nondegenerate limit.

\begin{acknowledgments}
This work was supported by the National Natural Science Foundation of
China under Grant No.\ 11334014, the National Basic Research Program of
China under Grant No.\ 2012CB922002 and the Strategic 
Priority Research Program of the
Chinese Academy of Sciences under Grant No.\ XDB01000000. 
\end{acknowledgments}


\begin{thebibliography}{0}
\bibitem{radisavljevic6} B. Radisavljevic, A. Radenovic, J. Brivio, V. Giacometti, and A. Kis, 
Nature Nanotech. {\bf 6}, 147 (2011).

\bibitem{yzhang12} Y. Zhang, J. Ye, Y. Matsuhashi, and Y. Iwasa, Nano Lett. {\bf
  12}, 1136 (2012).

\bibitem{qhwang7} Q. H. Wang, K. Kalantar-Zadeh, A. Kis, J. N. Coleman, and
  M. S. Strano, Nature Nanotech. {\bf 7}, 699 (2012).

\bibitem{cg25} A. Castellanos-Gomez, E. Cappelluti, R. Rold\'{a}n,
  N. Agra\"{i}t, F. Guinea, and G. Rubio-Bollinger, Adv. Mater. {\bf 25}, 899 (2013).



\bibitem{splendiani10} A. Splendiani, L. Sun, Y. Zhang, T. Li, J. Kim, C.-Y. Chim, G. Galli, and 
F. Wang, Nano Lett. {\bf 10}, 1271 (2010).

\bibitem{mak105} K. F. Mak, C. Lee, J. Hone, J. Shan, and T. F. Heinz, Phys. Rev. Lett. {\bf 105}, 136805 (2010).


\bibitem{eda11} G. Eda, H. Yamaguchi, D. Voiry, T. Fujita, M. Chen, and M. Chhowalla, Nano Lett. 
{\bf 11}, 5111 (2011).

\bibitem{korn99} T. Korn, S. Heydrich, M. Hirmer, J. Schmutzler, and C. Sch\"{u}ller, Appl. Phys. Lett. {\bf 99}, 
102109 (2011). 




\bibitem{cao3} T. Cao, G. Wang, W. Han, H. Ye, C. Zhu, J. Shi, Q. Niu, P. Tan, 
E. Wang, B. Liu, and J. Feng, Nature Commun. {\bf 3}, 887 (2012).


\bibitem{zeng7} H. Zeng, J. Dai, W. Yao, D. Xiao, and X. Cui, Nature Nanotech. {\bf 7}, 490 (2012).

\bibitem{mak7} K. F. Mak, K. He, J. Shan, and T. F. Heinz, Nature Nanotech. {\bf 7}, 494 (2012).

\bibitem{xiao108} D. Xiao, G.-B. Liu, W. Feng, X. Xu, and W. Yao, Phys. Rev. Lett. {\bf 108}, 
196802 (2012).


\bibitem{sallen86} G. Sallen, L. Bouet, X. Marie, G. Wang, C. R. Zhu, W. P. Han, Y. Lu, P. H. Tan, T. Amand, 
B. L. Liu, and B. Urbaszek, Phys. Rev. B {\bf 86}, 081301(R) (2012).


\bibitem{lagarde} D. Lagarde, L. Bouet, X. Marie, C. R. Zhu, B. L. Liu,
  T. Amand, P. H. Tan, and B. Urbaszek, Phys. Rev. Lett. {\bf 112}, 047401 (2014).

\bibitem{tao} T. Yu and M. W. Wu, arXiv:1401.0047.

\bibitem{gwang} G. Wang, L. Bouet, D. Lagarde, M. Vidal, A. Balocchi, T. Amand,
  X. Marie, and B. Urbaszek, arXiv:1402.6009.

\bibitem{glazov2} M. M. Glazov, T. Amand, X. Marie, D. Lagarde, L. Bouet, and
  B. Urbaszek, arXiv:1403.0108.


\bibitem{zhu84} Z. Y. Zhu, Y. C. Cheng, and U. Schwingenschl\"{o}gl, Phys. Rev. B {\bf 84}, 
153402 (2011).

\bibitem{chei} T. Cheiwchanchamnangij and W. R. L. Lambrecht, Phys. Rev. B {\bf 85}, 
205302 (2012).

\bibitem{kadantsev152} E. S. Kadantsev and P. Hawrylak, Solid State Commun. {\bf 152}, 909 (2012).

\bibitem{ochoa} H. Ochoa and R. Rold\'{a}n, Phys. Rev. B {\bf 87}, 245421 (2013).

\bibitem{hshi} H. Shi, H. Pan, Y.-W. Zhang, and B. I. Yakobson, Phys. Rev. B {\bf 87}, 155304 (2013).

\bibitem{rostami} H. Rostami, A. G. Moghaddam, and R. Asgari,
  Phys. Rev. B {\bf 88}, 085440 (2013).

\bibitem{zahid} F. Zahid, L. Liu, Y. Zhu, J. Wang, and H. Guo, AIP Advances
{\bf 3}, 052111 (2013).


\bibitem{kormanyos} A. Korm\'{a}nyos, V. Z\'{o}lyomi, N. D. Drummond, P. Rakyta, G. Burkard, and V. 
I. Fal'ko, Phys. Rev. B {\bf 88}, 045416 (2013).

\bibitem{cappelluti} E. Cappelluti, R. Rold\'{a}n, J. A. S.-Guill\'{e}n,
  P. Ordej\'{o}n, and F. Guinea, Phys. Rev. B {\bf 88}, 075409 (2013).


\bibitem{kosmider87} K. Ko\'{s}mider and J. F. Rossier, Phys. Rev. B {\bf 87}, 075451 (2013).

\bibitem{wangarXiv1} L. Wang and M. W. Wu, Phys. Lett. A, DOI: 
10.1016/j.physleta.2014.03.026.

\bibitem{wangarXiv2} L. Wang and M. W. Wu, Phys. Rev. B {\bf 89}, 115302 (2014).

\bibitem{nishikawa} Y. Nishikawa, A. Takeuchi, M. Yamaguchi, S. Muto, and
  O. Wada, IEEE J. Quantum Electron. {\bf 2}, 661 (1996).

\bibitem{kikkawa277} J. M. Kikkawa, I. P. Smorchkova, N. Samarth, and
  D. D. Awschalom, Science {\bf 277}, 1284 (1997).

\bibitem{kikkawa397} J. M. Kikkawa and D. D. Awschalom, Nature (London) {\bf
    397}, 139 (1999).

\bibitem{martin67} I. Martin, Phys. Rev. B {\bf 67}, 014421 (2003).

\bibitem{voros94} Z. V\"{o}r\"{o}s, R. Balili, D. W. Snoke, L. Pfeiffer, and
  K. West, Phys. Rev. Lett. {\bf 94}, 226401 (2005).

\bibitem{tombros} N. Tombros, C. Jozsa, M. Popinciuc, H. T. Jonkman, and
  B. J. van Wees, Nature (London) {\bf 448}, 571 (2007).

\bibitem{datta} S. Datta and B. Das, Appl. Phys. Lett. {\bf 56}, 665 (1990).

\bibitem{wunderlich} J. Wunderlich, B. G. Park, A. C. Irvine, L. P. Z\^{a}rbo,
  E. Rozkotov\'{a}, P. Nemec, V. Nov\'{a}k, J. Sinova, and T. Jungwirth, Science
  {\bf 330}, 1801 (2010).

\bibitem{koo} H. C. Koo, J. H. Kwon, J. Eom, J. Chang, S. H. Han, and
  M. Johnson, Science {\bf 325}, 1515 (2009).



\bibitem{dp} M. I. D'yakonov and V. I. Perel', Zh. Eksp. Teor. Fiz. {\bf 60}, 1954 (1971)
[Sov. Phys. JETP {\bf 33}, 1053 (1971)]; Fiz. Tverd. Tela (Leningrad) {\bf
    13}, 3581 (1971) [Sov. Phys. Solid State {\bf 13}, 3023 (1972)].

\bibitem{ey} Y. Yafet, Phys. Rev. {\bf 85}, 478 (1952); R. J. Elliot, {\it
    ibid.} {\bf 96}, 266 (1954).

\bibitem{wuning} M. W. Wu and C. Z. Ning, Eur. Phys. J. B {\bf 18}, 373 (2000); 
M. W. Wu, J. Phys. Soc. Jpn. {\bf 70}, 2195 (2001).


\bibitem{cheng101} J. L. Cheng and M. W. Wu, J. Appl. Phys. {\bf 101}, 073702
  (2007).

\bibitem{weng} M. Q. Weng and M. W. Wu, Phys. Rev. B {\bf 66}, 235109 (2002); 
J. Appl. Phys. {\bf 93}, 410 (2003).


\bibitem{peng84} P. Zhang and M. W. Wu, Phys. Rev. B {\bf 84}, 045304 (2011).

\bibitem{bishnoi} B. Bishnoi and B. Ghosh, J. Comput. Electron. DOI: 10.1007/s10825-013-0547-7.  

\bibitem{kormanyos2} A. Korm\'{a}nyos, V. Z\'{o}lyomi, N. D. Drummond, and
  G. Burkard, Phys. Rev. X {\bf 4}, 011034 (2014).

\bibitem{cheng75} J. L. Cheng, M. W. Wu, and I. C. da Cunha Lima, Phys. Rev. B
  {\bf 75}, 205328 (2007).

\bibitem{peng842} P. Zhang and M. W. Wu, Phys. Rev. B {\bf 84}, 014433 (2011).

\bibitem{peng79} P. Zhang and M. W. Wu, Phys. Rev. B {\bf 79}, 075303 (2009).


\bibitem{wureview} M. W. Wu, J. H. Jiang, and M. Q. Weng, Phys. Rep. {\bf 493},
  61 (2010).

\bibitem{yzhou} Y. Zhou and M. W. Wu, Phys. Rev. B {\bf 82}, 085304 (2010).

\bibitem{kaasbjerg} K. Kaasbjerg, K. S. Thygesen, and K. W. Jacobsen, Phys. Rev. B {\bf 85}, 
115317 (2012). 

\bibitem{cubic} It is noted that the cubic term in Eq.~(\ref{soctime}) is
  neglected here due to its marginal contribution to the spin diffusion
  contributed by the electron-impurity scattering as previously mentioned. This
  is very different from the case of the spin relaxation in the time domain in
  monolayer MoS$_2$.

\bibitem{novoselov102} K. S. Novoselov, D. Jiang, F. Schedin, T. J. Booth,
  V. V. Khotkevich, S. V. Morozov and A. K. Geim, PNAS {\bf 102}, 10451 (2005).

\bibitem{ayari101} A. Ayari, E. Cobas, O. Ogundadegbe, and M. S. Fuhrer,
  J. Appl. Phys. {\bf 101}, 014507 (2007).


\bibitem{mobility} With a typical electron density $N_e=10^{13}\ $cm$^{-2}$,\cite{radisavljevic6,novoselov102} 
the impurity density $N_i=4.4\times 10^{13}\ $cm$^{-2}$ is obtained by fitting
to the mobility $10\ $cm$^2$V$^{-1}$s$^{-1}$ at room temperature. 



\bibitem{sun151} B. Y. Sun and K. Shen, Solid State Commun. {\bf 151}, 1322
  (2011).

\bibitem{zhang112} P. Zhang, Y. Zhou, and M. W. Wu, J. Appl. Phys. {\bf 112},
  073709 (2012).

\bibitem{lwang87} L. Wang and M. W. Wu, Phys. Rev. B {\bf 87}, 205416 (2013).


\bibitem{highorder} The relative error between the inter-valley spin diffusion length 
with the orders $|l|\le 1$ and $|l|\le 2$ is about $18$~\% and the one between the
inter-valley spin diffusion length with the orders $|l|\le 2$ and
  $|l|\le 3$ is about $6$~\%.  


\bibitem{glazov} M. M. Glazov and E. L. Ivchenko, Zh. Eksp. Teor. Fiz. {\bf
    126}, 1465 (2004) [JETP {\bf 99}, 1279 (2004)].

\bibitem{magnetic} With the last term (i.e., the negligible one) in ${\bgreek\omega}^{\mu}$ [see Eq.~(\ref{socspatial})]
  neglected, the spin precession frequency in the spatial domain in monolayer
  MoS$_2$ is given by $(2\lambda_c+A_1k^2)m^*\mu\hat{\bf z}/(\hbar^2k_x)$, which
  is similar to the one provided by the external magnetic field\cite{peng79} [i.e.,
  $m^*g\mu_B{\bf B}/(\hbar^2k_x)$].   



\end{thebibliography}
\end{document}